# Present status of development of damping ring extraction kicker system for CLIC

Janne Holma[1], Mike Barnes[1], Caroline Belver-Aguilar[2], Angeles Faus-Golfe[2] and Fernando Toral[3]

1 – CERN – TE/ABT/FPS
CH-1211, Geneva 23 – Switzerland

2 – IFIC – CSIC-UV
Valencia – Spain
3 – CIEMAT
Madrid – Spain

The CLIC damping rings will produce ultra-low emittance beam, with high bunch charge, necessary for the luminosity performance of the collider. To limit the beam emittance blow-up due to oscillations, the pulse power modulators for the damping ring kickers must provide extremely flat, high-voltage pulses: specifications call for a 160 ns duration and a flattop of 12.5 kV, 250 A, with a combined ripple and droop of not more than ±0.02 %. The stripline design is also extremely challenging: the field for the damping ring kicker system must be homogenous to within ±0.01 % over a 1 mm radius, and low beam coupling impedance is required. The solid-state modulator, the inductive adder, is a very promising approach to meeting the demanding specifications for the field pulse ripple and droop. This paper describes the initial design of the inductive adder and the striplines of the kicker system.

## 1  Introduction

The CLIC design relies on the presence of Pre-Damping Rings (PDR) and Damping Rings (DR) to achieve the very low emittance, through synchrotron radiation damping, needed for the luminosity requirements of CLIC. To achieve high luminosity in the Interaction Point (IP), it is crucial that the beams have very low transverse emittance: the PDR and DR damp the beam to an extremely low emittance in all three dimensions. The PDR is required to decouple the wide aperture requirements of the incoming beams form the final emittance requirements of the main linac. The design parameters of the PDR and DR are dictated by target performance of the collider (e.g. luminosity), the injected beam characteristics or compatibility with the downstream system parameters: the emittances of the beams in the damping rings must be reduced by several orders of magnitude [1]. A schematic of the CLIC proposal is shown in Fig. 1.

Kickers are required to inject beam into and extract beam from the PDRs and DRs. Jitter in the magnitude of the kick waveform translates into beam jitter at the IP [1]. Thus the PDR and DR kickers, in particular the DR extraction kicker, must have a very small magnitude of jitter as well as low longitudinal and transverse beam coupling impedances. Table 1 shows the specifications for the PDR and DR kickers [2]: the specified stabilities include all sources of contributions such as ripple and droop. The values in Table 1 may be refined as the optics design progresses. Striplines have been chosen for the kicker elements in order to meet the specifications for the excellent field homogeneity and very low beam coupling impedances [3].

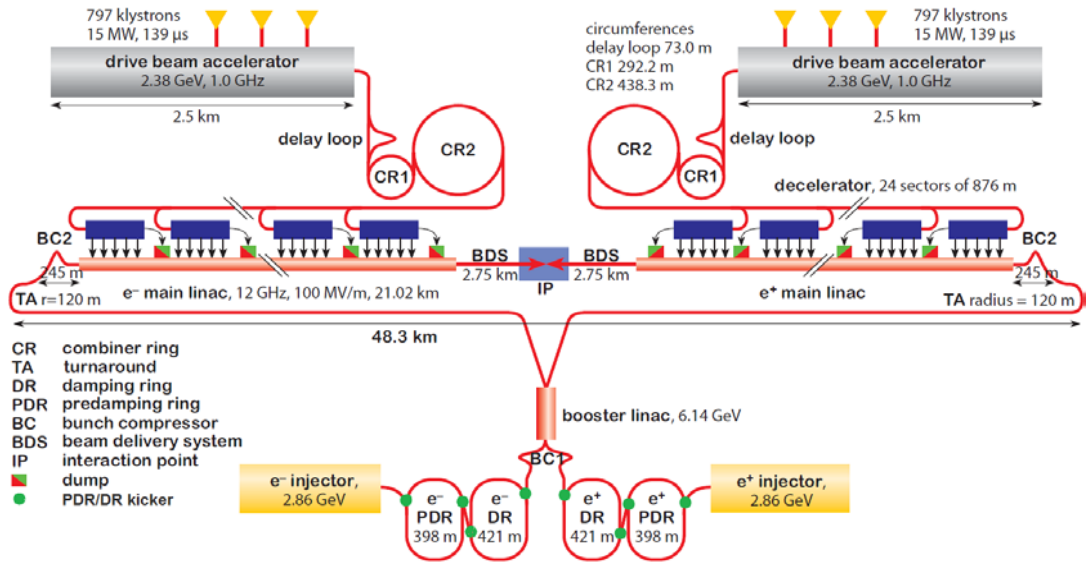

Figure 1: Proposed layout of CLIC facility.

| Parameter | PDR | DR |
|---|---|---|
| Field rise and fall times (ns) | 700 | 1000 |
| Pulse flat-top duration (ns) | 160 | 160 |
| Pulse flat-top reproducibility | ±1x10$^{-4}$ | ±1x10$^{-4}$ |
| Flat-top stability, at injection | ±2x10$^{-2}$ | ±2x10$^{-3}$ |
| Flat-top stability, at extraction | ±2x10$^{-3}$ | ±2x10$^{-4}$ |
| Injection field inhomogeneity (%) | ±0.1[A] | ±0.1[A] |
| Extraction field inhomogeneity (%) | ±0.1[A] | ±0.01[B] |
| Repetition rate (Hz) | 50 | 50 |
| Stripline and load impedance (Ω) | 50 | 50 |
| Pulse voltage per stripline (kV) | ±17 | ±12.5 |
| Stripline pulse current (A) | ±340 | ±250 |

[A]Over 3.5 mm radius   [B]Over 1 mm radius

Table 1: PDR and DR kicker specifications.

## 2  The stripline kicker system

Fig. 2 shows a simplified schematic of a stripline kicker system with an inductive adder. The two stripline plates are driven to an equal magnitude of voltage but of opposite polarity. The inductive adder feeds pulses towards the striplines (note: for simplicity,

LCWS11

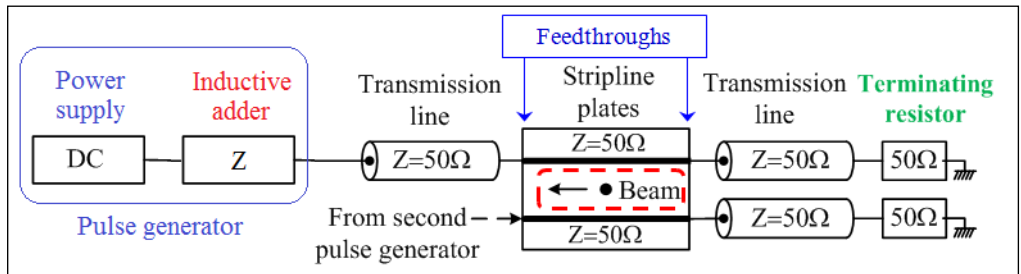

Figure 2: The stripline kicker system with an inductive adder

Fig. 2 only shows one inductive adder. For feeding both striplines, either one bipolar or two unipolar inductive adders are required). The pulse propagates through the striplines and is then deposited in a terminating resistor. The characteristic impedances of the inductive adder, transmission lines, striplines and terminating resistors are matched as far as possible to minimize reflections, which could cause ripple on the flat top of the deflection waveform: however it may not be feasible to match the odd mode characteristic impedance of the striplines to the characteristic impedance of the other elements of the kicker system [4].

To limit the beam emittance blow-up due to oscillations, the pulse power modulators for the DR kickers must provide extremely flat, high-voltage, pulses: specifications call for a 160 ns duration flattop of 12.5 kV, 250 A, with a combined ripple and droop of not more than ±0.02% (Table 1). Fig. 3 shows the definition of the pulse required for the CLIC DR & PDR.

- ➢ Rise time: is the time needed to reach the required voltage (including settling time);
- ➢ Settling time: is the time needed to damp oscillations to within specification;
- ➢ Droop & ripple: time window during which the combined droop and ripple must be within specification;
- ➢ Reproducibility: maximum difference allowed between two consecutive pulses.
- ➢ Fall time: time for voltage to return to zero.

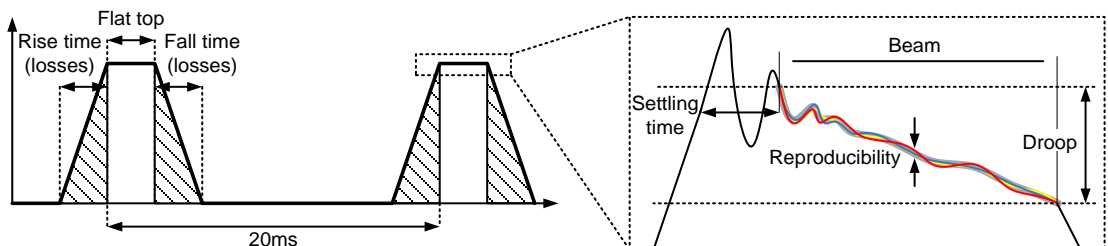

Figure 3: Definition of the pulse required for the CLIC PDR and DR kickers.

The possible sources of ripple, droop and irreproducibility of the deflection waveform include:
- ➢ Attenuation and dispersion in the transmission lines
- ➢ Feedthroughs
- ➢ Striplines
- ➢ Terminating resistor (frequency dependence of value, long-term stability and temperature will affect ripple and reproducibility of waveform)
- ➢ Non-ideal impedance matching of the system.

Several of these items have been discussed in details in ref. [4].

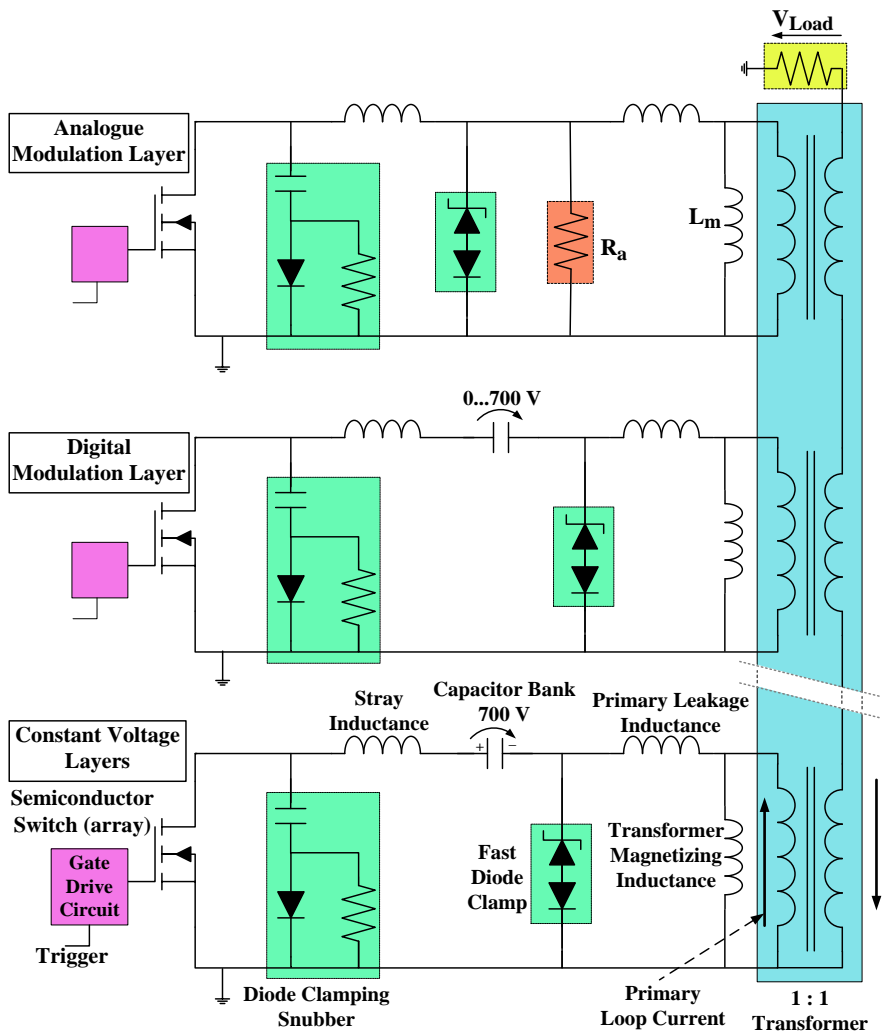

Figure 4: Schematic of an inductive adder with a constant voltage layer, a digital modulation layer and an analogue modulation layer.

# 3 The inductive adder

A review of literature of existing pulse generators has been carried out and an inductive adder (Fig. 4) has been selected as a very promising means of achieving the specifications for the PDR and DR extraction kickers [5]. The inductive adder is a solid-state modulator, capable of providing relatively short and precise pulses. With a proper design of the adder it may be possible to meet the ripple and droop requirements of the PDR kicker [6]. However, digital and analogue modulation of the output pulse may be required to reach the specifications of the DR kicker. A method to measure the output ripple of the pulse generator will also require significant development [7].

The inductive adder concept has many advantages with respect to other topologies [8]: the control electronics of the switches can be referenced to ground and there are no electronics referenced directly to the high voltage output pulse. If one layer fails to switch on, the output voltage is still a significant proportion of the required magnitude. The same stack can be used to generate positive, negative or bipolar pulses, depending on the grounding of the secondary winding. The primary impedance of the adder is low, which helps to minimize the sum of the primary layer voltages, for a given load voltage. In addition, the same adder design can probably be used for both the CLIC PDR and DR kicker modulators.

The disadvantages of the adder are that it is complex requiring many components – although this is good for redundancy – and each primary layer must switch full load current. The fault current may also be very large. In general, the pulse width of the adder is limited by saturation of the transformer core. The layers of the inductive adder must be designed to have very small loop inductances, which may require custom-made capacitors.

## 3.1 Selection of the main components for the inductive adder

### 3.1.1 *Semiconductor switches*

MOSFETs have been selected as the first candidate switch for the PDR and DR kickers, because they have many advantages over other switches. Switching transitions of MOSFETs are very fast, which allows short duration pulses. In addition MOSFETs are easy to control and they have low time jitter, usually in the order of sub-nanoseconds. Fast power MOSFETs are also available for operation in their linear region, which is required if analogue modulation techniques are used.

The operating voltage of the switches defines the number of layers in the adder stack. MOSFETs are available with voltage rating of up to 1.2 kV: for long-term reliability, while minimizing the number of layers, the solid-state switches will be operated at 60% to 70% of their voltage rating. Implementing the 12.5 kV DR kicker modulator with switches with an operating voltage of 700 V will require a minimum of 18 layers.

### 3.1.2 *Capacitors*

The layers of the inductive adder must be designed to have very small loop inductances, which may require custom-made capacitors. The required capacitance per layer, to directly achieve a given load voltage droop, without using modulation techniques, is proportional to the number of layers [5]. For a load with 160 ns wide flattop pulses of 12.5 kV and 250 A, allowing 0.02% (2.5 V) for droop requirements (the other 0.02% is an allowance for ripple), and an inductive adder consisting of 18 layers, the required capacitance per layer is 320 µF. This includes an allowance for a magnetizing current of 30 A. However modulation techniques should permit a lower value of capacitor to be used [6].

*3.1.3   Transformer cores*

The maximum output pulse-width of the inductive adder is limited by the saturation of the transformer core. The minimum required cross-sectional area (CSA) of the core is inversely proportional to the available flux-density swing [5]: the core can be biased to increase the available swing and therefore reduce the required CSA.
To minimize load voltage rise-time, eddy currents in the cores must be controlled: tape-wound cores must have adequate insulation to avoid breakdown between magnetic laminations. For a toroidal core it can be shown that the interlaminar voltage, $V_{il}$, is given by [5]:

$$V_{il} = \frac{A_{il}}{2 \cdot \pi \cdot r} \cdot N \cdot \mu_0 \cdot \left( i \cdot \frac{d\mu_r}{dt} + \mu_r \cdot \frac{di}{dt} \right)$$

where $A_{il}$ is the interlaminar cross-sectional area, $r$ is the radius of interest, $N$ is the number of turns and $i$ is current. For a non-magnetic interlaminar insulation, the first term in parentheses is equal to zero. The above equation shows that the interlaminar voltage is highest near to the inside radius of the core.
Temperature effects can have a significant influence upon the magnetic parameters of the transformer core and this must be carefully considered at the design stage [9].

*3.1.4   Dimensioning of stack secondary winding*

The secondary winding of the adder stack will be a single turn conducting rod. The output impedance of the stack will be matched to the impedance of the transmission lines and terminating resistor (Fig. 2) to minimize reflections. The primary leakage inductance of the transformer must be taken into consideration in defining the dimensions of the secondary conductor and hence its inductance and capacitance. Equations for approximately dimensioning the inductive adder cell have been published in ref. [10].

*3.1.5   Design steps*

The design of the dimensions of the inductive adder is an iterative process. The process can be carried out as follows: (a) choose the type of semiconductor switches, their voltage rating and operating voltage; (b) calculate the number of primary layers, including redundancy; (c) estimate the value of capacitance per layer; (d) determine the minimum CSA of the magnetic core – including a suitable safety margin; (e) determine the dimensions of the coaxial transformer structure. Other design constraints and more detailed presentation about design steps have been published in refs. [6, 11].

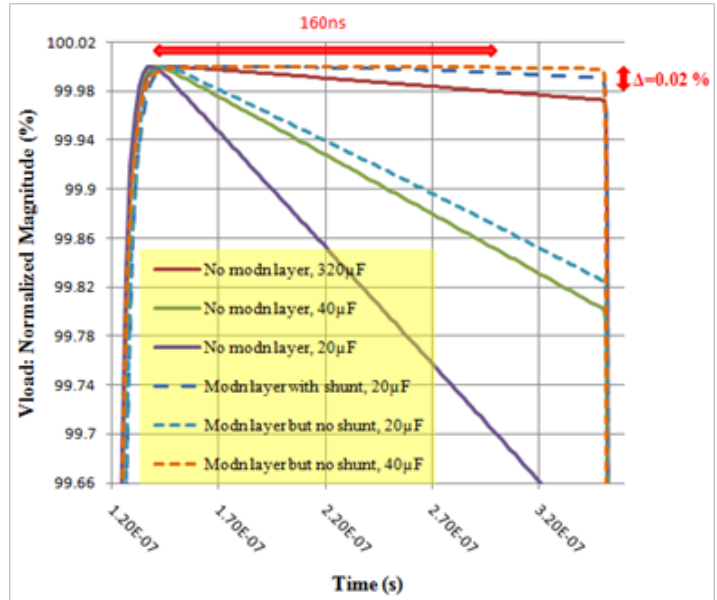

Figure 5: Simulation results for an inductive adder with different capacitances, with and without an analogue modulation layer.

### 3.2 Compensation of droop and ripple

The ripple and droop requirement of the DR kicker is tighter than for any pulse power modulator found in the literature and various techniques are being studied to achieve the specifications. Fig. 5 shows simulation results with different capacitances per layer and with a passive and active analogue modulation layer. Without modulation 320 µF per layer is required to achieve the specified droop: with analogue modulation, the droop is within specification with a capacitance per layer of 40 µF [5]. For the prototype inductive adder it is proposed to use 80 µF per layer. A feed-forward compensation of known or predictable ripple components could also be applied with active droop cancellation: this will be investigated.

## 4 Stripline design

The demanding specifications for the CLIC PDR and DR kickers are shown in Table 1. The specifications for the PDR and DR stripline kickers include low longitudinal and transverse beam coupling impedances, high stability and reproducibility of the field, excellent field homogeneity and ultra-high vacuum. The DR extraction kicker has the most demanding specification for field homogeneity. A stripline kicker consists of two parallel metallic electrodes which are connected at each end to the external circuit by coaxial feedthroughs. Each stripline is driven to equal but opposite potential. An example of stripline kicker is shown in fig. 6.

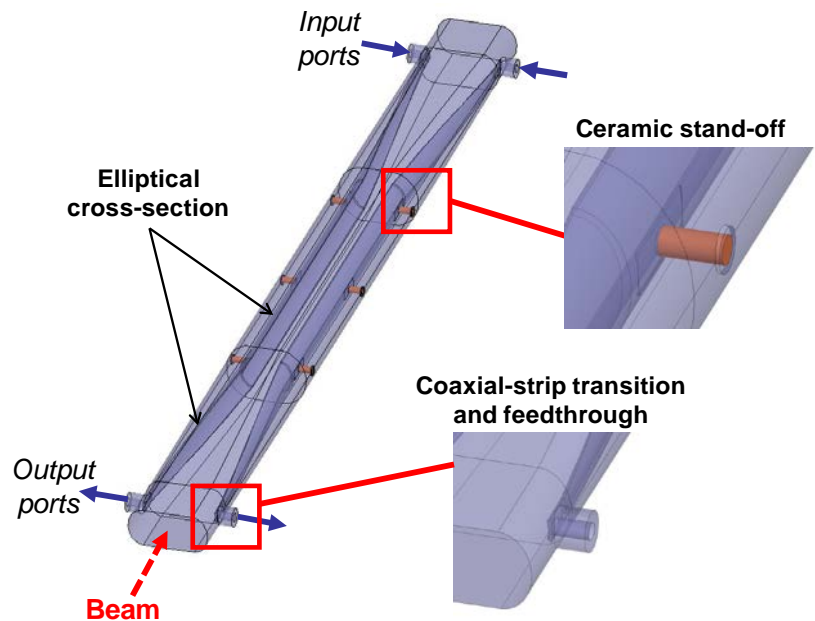

Figure 6: DAΦNE stripline kicker [13].

The kicker cross-section defines the characteristic impedance of the striplines and the homogeneity of the field inside the aperture. The cross-section must be optimized to achieve the desired characteristics impedance and the required field homogeneity [12, 14]. The beam impedance defines the interaction of the beam with the kicker, resulting in beam energy loss and a beam shape perturbation: the beam impedance is dependent upon the even-mode characteristic impedance [4, 14]. The beam impedance is desired to be low to avoid beam instabilities. The beam impedance can be reduced by tapering of electrodes, which means that the distance between the electrodes and the beam pipe is reduced in order to smooth the change of geometry seen by the beam (Fig. 6), when it passes through the kicker aperture. Optimization studies of the DR kicker striplines have been presented in refs. [12, 14]. Fig. 7 shows the sensitivity of the field homogeneity to the parameter variations from an optimized design of the DR kicker. Fig. 8 and 9 show the contour plots of the field homogeneity in the kicker aperture for an optimized design. According to the contour plots, the optimized design meets the tight field homogeneity specifications, corresponding to ±0.1 % for the DR injection kicker over 3.5 mm of radius and ±0.01 % over the 1 mm of radius of the aperture.

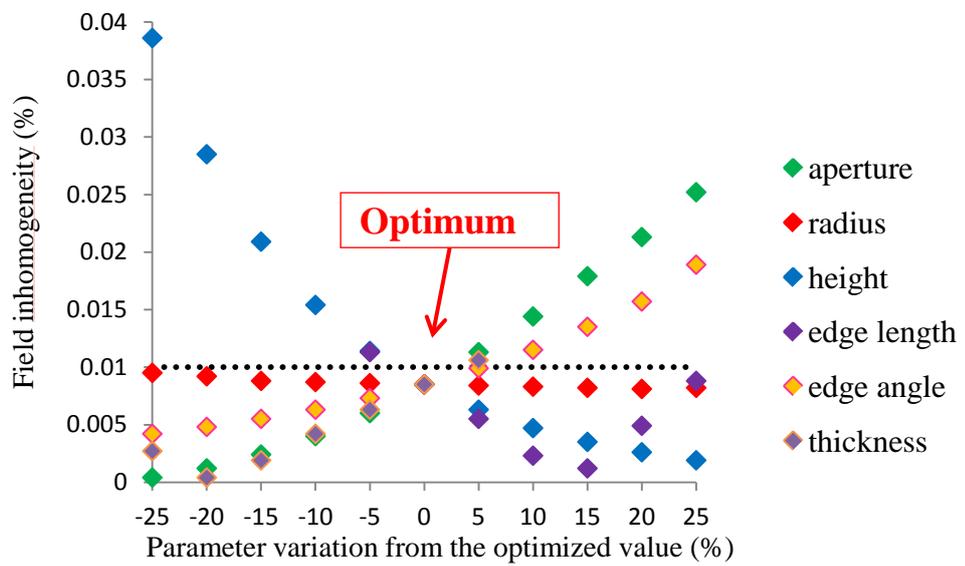

Figure 7: The sensitivity of the field homogeneity to parameter variations from an optimized design [14].

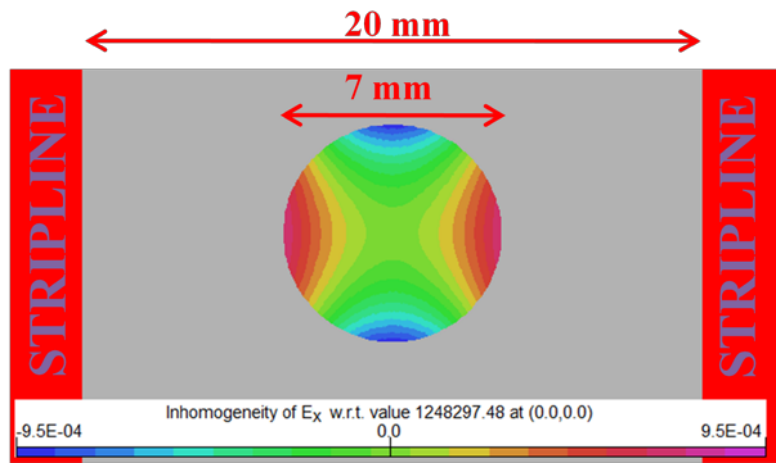

Figure 8: Field homogeneity in the kicker aperture, over 3.5 mm radius, for an optimized design.

LCWS11

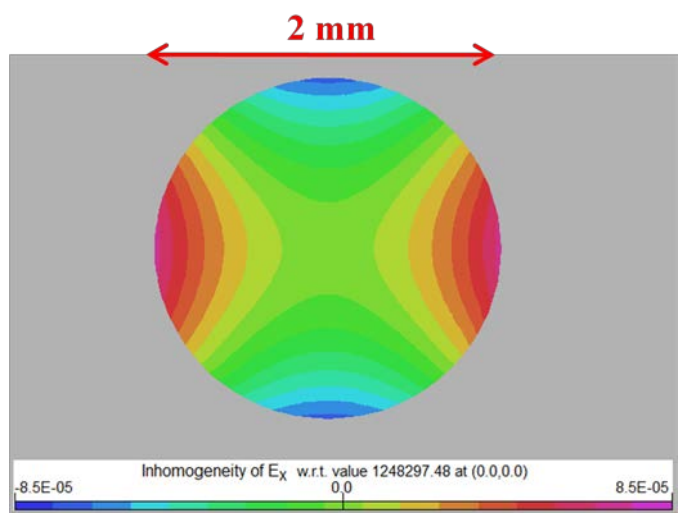

Figure 9: Field homogeneity in the kicker aperture, over 1 mm radius, for the optimized design.

## 5 Conclusion

The DR kicker system for CLIC is challenging due to its requirement of excellent field homogeneity, low beam coupling impedance and ultra-low ripple and droop. Striplines with suitable tapers can achieve the required transverse and longitudinal beam coupling impedance. A set of striplines will be prototyped under the Spanish Science for Industry Program: the prototype striplines are scheduled to be built in summer 2012.

An inductive adder is being investigated for the pulse generators for the PDRs and DRs.. The inductive adder design is such as to permit pulse shaping and is expected to be inherently highly reliable and provide good machine protection. The preliminary design of the prototype inductive adder has been carried out and tests of the candidate components have commenced. The first prototype adder is scheduled to be built by the end of 2012.

## 6 References

[1] Y. Papaphilippou, "CLIC Damping Ring Beam Transfer Systems" (2010).

[2] Y. Papaphilippou, "Parameter Specification, EDMS#989080, Kickers for the CLIC Damping and Predamping Rings", PBS reference: 1.2._.10.

[3] M.J. Barnes et al, "CLIC Pre-Damping and Damping Ring Kickers: Initial Ideas to Achieve Stability Requirements", Proc. IPAC'10 (2010), pp3305-3307.

[4] M.J. Barnes et al, "CLIC Beam Transfer System", CLIC CDR, to be publ.

[5] J. Holma et al, "Preliminary Design of a Pulse Power Modulator for the CLIC DR Extraction System", to be publ., Proc. 18$^{th}$ IEEE PPC (2011).

[6] J. Holma, M.J. Barnes, "Pulse Power Generator Development for the CLIC Damping Ring Kickers",

CERN TE Note, to be publ (2012).

[7] M.J. Barnes, J. Holma, "An Inductive Adder as a Low-Jitter, Ultra-Flat, DR Extraction Pulser", LεR2011, Greece (2011). http://indico.cern.ch/getFile.py/access?contribId=49&sessionId=22&resId=3&materialId=slides&confId=148596

[8] E.G. Cook et al., "Solid-State Modulator R&D at LLNL," Int. Workshop on Recent Progress of Induction Accelerators, Tsukuba, Japan (2002).

[9] C. Jensen, private communication (2011).

[10] C. Burkhart et al, "MIA Modulator for the LLNL 'Small Recirculator' Experiment", Nuclear Instruments and Methods in Physics Research Section A: Accelerators, Spectrometers, Detectors and Associated Equipment, Vol. 464, No. 1-3 (2001).

[11] J. Holma, M.J. Barnes, "Preliminary Design of an Inductive Adder for CLIC Damping Rings", Proc. IPAC'11 (2011), pp3409-3411.

[12] C. Belver-Aguilar et al, "Striplines for CLIC Pre-Damping and Damping Rings", Proc. IPAC'11 (2011), pp1012-1014.

[13] D. Alesini, "Fast RF Kicker Design", ICFA Mini-Workshop on Deflecting/Crabbing Cavity Applications in Accelerators, Shanghai (2008).

[14] C. Belver-Aguilar, "R&D on Striplines for the CLIC DR Kickers", ICFA Beam Dynamics Mini Workshop on Low Emittance Rings (2011), http://indico.cern.ch/getFile.py/access?contribId=50&sessionId=22&resId=1&materialId=slides&confId=148596.